\documentclass[final,3p,times,twocolumn]{elsarticle}

\usepackage{graphics}

\usepackage{amssymb}
\usepackage{bm}
\usepackage{amsmath}

\usepackage{color}
\usepackage[latin1]{inputenc}
\usepackage[english]{babel}
\usepackage{dsfont}

\journal{Measurment}

\newcommand{\NA}{{N_{\rm A}}}
\newcommand{\eSi}{{^{28}{\rm Si}}}
\newcommand{\rE}{{\rm E}}
\newcommand{\rH}{{\rm H}}
\newcommand{\bsigma}{{\boldsymbol{\sigma}}}
\newcommand{\bx}{{\boldsymbol{x}}}
\newcommand{\bh}{{\boldsymbol{h}}}
\newcommand{\rmd}{{\rm d}}
\newcommand{\rme}{{\rm e}}

\def\app#1#2{%
  \mathrel{%
    \setbox0=\hbox{$#1\sim$}%
    \setbox2=\hbox{%
      \rlap{\hbox{$#1\propto$}}%
      \lower1.1\ht0\box0%
    }%
    \raise0.25\ht2\box2%
  }%
}

\begin{document}

\begin{frontmatter}



\title{Model uncertainty and reference value of the Planck constant}


\author{G.\ Mana\corref{1}}
\ead{g.mana@inrim.it}
\cortext[1]{Corresponding author}
\address{INRIM -- Istituto Nazionale di Ricerca Metrologica, str.\ delle Cacce 91, 10135 Torino, Italy}

\begin{abstract}
Statistical parametric models are proposed to explain the values of the Planck constant obtained by comparing electrical and mechanical powers and by counting atoms in $\eSi$ enriched crystals. They assume that uncertainty contributions -- having heterogeneous, datum-specific, variances -- might not be included in the error budgets of some of the measured values. Model selection and model averaging are used to investigate data consistency, to identify a reference value of the Planck constant, and to include the model uncertainty in the error budget.
\end{abstract}


\end{frontmatter}


\section{Introduction}
The definition of a system of units on the basis of conventional values of fundamental constant of physics \cite{Milton:2014} is motivating efforts on determinations of the Planck constant \cite{Massa:2012}. The most accurate data come from the comparison of mechanical and electrical powers by watt-balance experiments \cite{Stock:2013,Steiner:2013} and the count of the atoms in $\eSi$ enriched silicon balls \cite{Bettin:2013}. Four $h$ determinations comply with the accuracy required to make the kilogram redefinition feasible \cite{Andreas2:2011,Azuma:2015,Schlamminger:2015,Sanchez:2014,Wood:2015}. A statistical analysis of these results is necessary to check their consistency and to chose a reference value of the Planck constant.

Data analysis is usually carried out by selecting a model and by processing the measurement results as if they had generated by it. This approach ignores the model uncertainty and can lead to underestimates of the uncertainty, to overconfident inferences, and to decisions that are more risky than one thinks they are. Questions are: How accurately does a model explain the data and what is the impact of the model uncertainty on the measurand estimate and the inferences that we draw from the measurement results? Given an uncertain data model and a measurand estimate based on it, how can the total uncertainty of the measurand value be assessed?

Probability calculus can select the model most supported by the data and include the uncertainty into the analysis and uncertainty budget \cite{Linden:2014}; an example investigating the choice of the degree when fitting a polynomial to noisy data is given in \cite{Lago:2014}. The choice of a measurand value from inconsistent data-sets is investigated in \cite{Dose:2007,Elster:2010,Toman:2012,Mana:2012}.

This paper builds on these works and delivers some additional results. Firstly, it considers models where the standard deviations of a data subset -- the empty set and the whole data set included -- might be larger than the associated uncertainties; but, we do not know what this subset is. Secondly, it chooses the uninformative prior distribution of the unknown standard-deviations by requiring that Gaussian sampling-distributions of the measurement results are equiprobable. A novelty is that, if these standard-deviations are not of interest, marginalization allows an analytical expression of the measurement-result distributions to be given, no matter what the standards deviations -- greater than or equal to the associated uncertainties -- may be. Eventually, since one of the subset does apply, this paper tests the data consistency by comparing the probability of each subset is the right one given the data and suggests a reference value of the Planck constant by averaging over all the subsets. In this way, all the data determine the reference value, no measurement result is excluded, and none is considered fully reliable or suspicious.

\section{Planck constant values}
The starting point of the analysis is the list in table \ref{T1}. In 2014, the Bureau International des Poids et Mesures (BIPM) carried out a campaign of mass calibration with respect to the international prototype, in anticipation of the redefinition of the kilogram \cite{Stock:2015}. This brought to light an offset of the BIPM as-maintained mass unit, which was traceable to the prototype in 1992. Therefore, the mass values used in the watt-balance and atom counting experiments, were suitably corrected.

The IAC's $\NA$ values are converted into Planck constant values via the molar Planck constant $\NA h = 3.9903127176(28)\times 10^{-10}$ Js mol$^{-1}$, which has a negligible uncertainty \cite{Mohr:2010}. The correlation of the $\NA$ values reported in 2011 and 2015 by the IAC is investigated in \cite{Borys:2015}, which gives also the mean of the correlated values. To avoid complications due to the correlation, the input datum for this analysis is the mean of the 2011 and 2015 IAC's values.

\begin{figure}\centering
\includegraphics[width=7.5cm]{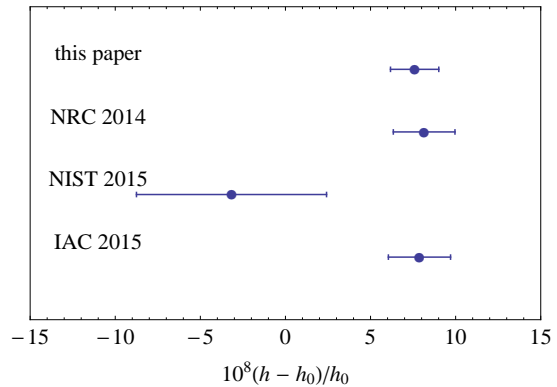}
\caption{Measured values of the Planck constant; $h_0 = 6.62606957\times 10^{-34}$ J s is the value recommended by the Committee on Data for Science and Technology.} \label{h:data}
\end{figure}

The values selected for this analysis are labelled from 1 to 3 in table \ref{T1}; they are shown in Fig.\ \ref{h:data}. The BIPM estimated the calibration uncertainty as 3 $\mu$g; this uncertainty affects all the mass values in the watt-balance and atom counting experiments. The table \ref{T1} gives the fractions of this systematic component of the uncertainty budget; the correlation of any pair of $h$ values can be obtained by multiplying the pair's systematic fractions.

\renewcommand\tabcolsep{6pt}
\begin{table}\centering\footnotesize
\caption{\label{T1}Measured values of the Planck constant; $f$ is the fraction of the systematic contribution to the uncertainty budget.}
\begin{tabular}{llllll}
  \hline
Lab &year &reference &label &$10^{34} h$ / Js &$f$ \\
  \hline
IAC$^a$  &2011 &\cite{Andreas2:2011,Azuma:2015}          &- &6.62606991(20) \\
IAC$^a$  &2015 &\cite{Azuma:2015}                        &- &6.62607016(13) \\
IAC$^b$  &2015 &\cite{Borys:2015}                        &1 &6.62607009(12) &0.16\\
NIST     &2015 &\cite{Schlamminger:2015}                 &2 &6.62606936(37) &0.05\\
NRC      &2014 &\cite{Sanchez:2014,Wood:2015,Azuma:2015} &3 &6.62607011(12) &0.17\\
         &     &this paper                               &- &6.626070073(94) \\
  \hline                                                \\
\multicolumn{5}{l}{IAC -- International Avogadro Coordination}\\
\multicolumn{5}{l}{NIST -- National Institute of Standards and Technology (USA)}\\
\multicolumn{5}{l}{NRC -- National Reasearch Council (Canada)} \\
\multicolumn{5}{l}{$^a$these values' correlation is 17\% \cite{Borys:2015}}\\
\multicolumn{5}{l}{$^b$average of the 2011 and 2015 IAC's correlated-values}\\
\end{tabular}
\end{table}

In 2012, the consultative committee for mass and related quantities of the International Committee for Weights and Measures recommended that "{\it ... the values provided by the different experiments be consistent at the 95\% level of confidence"} \cite{CCM:2012}. Since the confidence level is a concept associated to the Neyman's confidence interval \cite{Neyman:1937,Mana:2014}, the meaning of this recommendation is not very clear.

A way to examine the data consistency might be the significance test of Fisher \cite{Linden:2014}. Assuming that the data are independent normal-variables having the same mean $h$ and standard deviations $\sigma_i$ equal to the associated uncertainties, $u_i$ -- which is the consistent-data or null hypothesis, a test statistic is the Pearson $\chi^2$ variable \cite{Pearson:1900}. By choosing a 5\% significance level, the expected 95\% quantile is $\chi^2_{0.95} = 6.0$. For this data set, the observed value is $\chi^2 = 3.8$; since this value is less than the $\chi^2_{0.95}$ rejection level, the consistent-data hypothesis is accepted. The test ensures that the probability of rejecting the consistent-data model when it is true is 5\%, but to accept the consistent-data hypothesis as correct is an {\it argumentum ad ignorantiam} fallacy.

In order to assess the data consistency, we must calculate the probability of the null hypothesis; this requires to include it into a wider hypothesis space. To this end, we consider underestimations of the data uncertainties. Accordingly, each datum $x_i$ is thought to be a random variable having mean $h$ and variance $\sigma_i^2 = u_i^2+\lambda_i^2$, where, when reporting the measurement uncertainty, a datum-specific contribution to the variance, $\lambda_i^2$, was omitted. It is also possible that some measurement uncertainty was correctly evaluated -- that is, for these measurements, $\sigma_i=u_i$. Of course, all the measurement uncertainty might be correctly evaluated.

Our assumption is that there exists a subset of good data -- which might be empty set or the full data list -- having correct uncertainty assessments; its elements $x_i$ are independent realizations of random variables having variances $u_i^2$. For the remaining data, the uncertainties $u_i$ are only lower bounds to the standard deviations, that are additional model parameters. The good data can not be predetermined; instead, all the subsets will be taken in turn as the sought good-data subset. The final $h$ estimate will be obtained by model averaging using the probability of the each subset being the good one.

\section{Theoretical framework}
Before going into the specific application to the data in the table \ref{T1}, this section outlines the theoretical framework of the analysis.

\subsection{Model selection}
In order to explain the measurement results, we consider a number of parametric statistical models -- say, $M_A, M_B, ...$ -- where each model is parameterized by the measurand $h$ and, possibly, a set of nuisance parameters $\bsigma$. We assume that the set's models are mutually exclusive and complete, that is, $\dot{\bigvee}_n M_n = {\rm True}$. Given the measurement results $\bx=[x_1, x_2, x_3]^{\rm T}$ and the data likelihood $L(\bx|h,\bsigma, M)$, one proceeds by assigning a prior probability distribution $\pi(h,\bsigma|M)$ to the model parameters and a prior probability $\Pi(M)$ to each model. Next, by using the product rule of the probabilities, the joint distribution of the data, parameters, and models is
\begin{equation}\label{product}
 P(\bx, h,\bsigma, M) = L(\bx|h,\bsigma, M) \pi(h,\bsigma|M) \Pi(M) .
\end{equation}
According this hierarchical model, firstly, $M$ is sampled from $\Pi(M)$; then, the model parameters $h$ and $\bsigma$ are sampled from $\pi(h,\bsigma|M)$; eventually, the data $\bx$ are sampled from $L(\bx|h,\bsigma, M)$.

Through conditioning and marginalization, $P(\bx,h,\bsigma, M)$ can be used to obtain the post-data distributions of interest. By conditioning (\ref{product}) on $\bx$ and $M$, one gets the post-data probability distribution of the parameters given the model and data,
\begin{equation}\label{post}
 P(h,\bsigma|\bx, M) = \frac{L(\bx|h,\bsigma, M) \pi(h,\bsigma|M)}{Z(\bx|M)} ,
\end{equation}
where the normalizing factor
\begin{equation}\label{evidence}
 Z(\bx|M) = \int_{-\infty}^{+\infty} {\rm d}h \int_\Sigma L(\bx|h,\bsigma, M) \pi(h,\bsigma|M)\, {\rm d}\bsigma
\end{equation}
is the data evidence and the integration is carried out over the parameter space.

The marginalization of (\ref{product}) over the model parameter and the conditioning on the data yields the model probability given the data,
\begin{equation}\label{model}
{\rm Prob}(M|\bx) = \frac{Z(\bx|M) \Pi(M)}{\sum_M Z(\bx|M) \Pi(M)} .
\end{equation}
Within this framework, ${\rm Prob}(M|\bx)$ is the updated probability that $M$ is the model sampled in the first step of (\ref{product}). The pre-data distributions $\pi(h,\bsigma|M)$ and $\Pi(M)$ synthesize the uncertainty before the measurements are carried out; subsequently, the updated distributions $P(h,\bsigma|\bx, M)$ and ${\rm Prob}(M|\bx)$ synthesize the uncertainty after the data $\bx$ have been observed.

\subsection{Model averaging}
The simplest way to select a model is to choose the most probable. When no single model stands out, the expression of the uncertainty may require to report a set of models along with their probabilities. Model averaging is an alternative that incorporates model uncertainty. After marginalization of (\ref{product}) over the models and nuisance parameters $\bsigma$ and conditioning on the data, the distribution of the $h$ values is
\begin{equation}\label{measurand}
 P(h|\bx) = \sum_M P(h|\bx, M)  {\rm Prob}(M|\bx) ,
\end{equation}
where
\begin{eqnarray}\label{marginal}
 P(h|\bx, M) &= &\int_\Sigma P(h,\bsigma|\bx, M)\, {\rm d}\bsigma \nonumber \\
             &= &\frac{\displaystyle\int_\Sigma L(\bx|h,\bsigma, M) \pi(h,\bsigma|M)\, {\rm d}\bsigma}{Z(\bx|M)} .
\end{eqnarray}
By averaging over the models, (\ref{measurand}) incorporates the model uncertainty embedded in $\Pi(M)$. A point estimate of $h$ is the mean of (\ref{measurand}). Hence,
\begin{equation}\label{mean}
 \rE(h|\bx) = \sum_M \rE(h|\bx,M) {\rm Prob}(M|\bx) ,
\end{equation}
where $\rE(h|\bx,M)$ is the mean of (\ref{marginal}).

\section{Random effect model}
To explain the data, our hypothesis is as follows: the measured values $x_i$ of the Planck constant are independently sampled from distributions having the same mean $h$ and different variances. By maximizing the Shannon entropy, this information is synthesised by Gaussian sampling distributions, that is, $x_i \sim N(x|h,\sigma_i)$. Both $h$ and $\sigma_i$ are unknown, but the uncertainties $u_i$ associated to the data are lower bounds for $\sigma_i$, that is, $\sigma_i \ge u_i$. In this way, we allow for unknown errors that are datum-specific and have not been included in the uncertainty budgets of the measured values. We do not assume the existence of these errors: For some datum -- may be none, may be all -- the $\sigma_i = u_i$ identity might hold. Therefore, our hypothesis space contains as many models as the number of subsets of the measured values -- the empty set and the input data included, where each subset identifies the measurements whose associated uncertainty is the standard deviation, that is, $\sigma_i = u_i$.

The measurements are assumed uncorrelated, which is not exactly true. Though it is possible to include correlations \cite{Bodnar:2015}, this is beyond the scope of this analysis.

\subsection{Data likelihood}
Let us consider any measured value. If the associated uncertainty $u$ is the standard deviation, its sampling distribution is $N(x|h,u)$. Contrary, if the associated uncertainty is a lower bound for the standard deviation, the sampling distribution is $N(x|h,\sigma)$, where $\sigma \ge u$; in both cases, we omitted the $i$ subscript. In the latter case, the unknown variance $\sigma^2$ is a nuisance parameter that will be eliminated by the marginalization (\ref{marginal}). The data likelihood is
\begin{equation}\label{lik}
 L(\bx|h,\bsigma,M_A) = \prod_{i\in A, j\in \bar{A}} N(x_i|h,u_i) N(x_j|h,\sigma_j) ,
\end{equation}
where $A$ is a subset of good data $\{ x_i| x_i \sim N(x|h,u_i) \}$, $\bar{A}$ is the complement of $A$ in the hypothesis space, that is, $\bar{A} = \{x_j| x_j \sim N(x|h,\sigma_j)\}$, and $\bsigma =[\sigma_j| j\in\bar{A}]^{\rm T}$.

\subsection{Prior distributions}
From a theoretical viewpoint, when assigning probabilities to the values of $h$, it is impossible to get rid of the prior distribution $\pi(h,\bsigma)$. This distribution must synthesize the pre-data knowledge about $h$ and $\bsigma$. Ignorance means that the sampling distributions $L(\bx|h,\bsigma,M_A)$ must be equiprobable. Since they form a Riemannian manifold -- whose natural metric is the Fisher information $\mathbb{J}(h,\bsigma)$ \cite{Linden:2014,Rodriguez:1999,Amari:2007}, the distribution of the $(h,\bsigma)$ coordinates is proportional to the volume element $\sqrt{\det(\mathbb{J})}\,\rmd h\,\rmd\bsigma$. After normalisation, one gets the Jeffreys prior
\begin{equation}\label{prior}
 \pi(h,\bsigma) = \frac{1}{V_0} \prod_{j\in \bar{A}}\frac{u_j}{\sigma_j^2} ,
\end{equation}
where $V_0$ is the volume of the $h$ subspace and $\sigma_j > u_j$.

The relevance of (\ref{prior}) resides in making (\ref{post}) consistent {\it vs.} the transformations of the $h$ and $\bsigma$ variables. In fact, if we transform $h$ and $\bsigma$, the left-hand side of (\ref{post}) transforms according to the usual change-of-variable rule. What happens to the right-hand side is that the transformation Jacobian combines with $\mathbb{J}(h,\bsigma)$ to give the Fisher information about the new variables.

For example, if we consider a single datum and reparameterise the sampling distribution by $(h,\sigma^2)$ -- which corresponds to a coordinate change in the $N(x|h,\sigma)$ manifold -- the volume element changes from $\sqrt{2}\,\rmd h\,\rmd \sigma/\sigma^2$ to $\rmd h\,\rmd \sigma^2/(\sqrt{2}\sigma^3)$. As regards the pre-data distribution, by applying the distribution transformation rule, it changes from $u/(V_0\sigma^2)$ to $u/(2V_0\sigma^3)$, which are both consistent with $\pi(h,\sigma)\propto \sqrt{\det(\mathbb{J})}$.

\subsection{Marginalisation}
The integral (\ref{marginal}) reduces to the calculation of the sampling distribution of the measurement result $x$, given the Planck constant and the variance lower bound $u^2$,
\begin{equation}\label{sampling:2}
 G(x|h,u) = \int_u^{+\infty} \frac{u N(x|h,\sigma)}{\sigma^2} \, \rmd \sigma =
 \frac{ \left[ 1-\rme^{ -\frac{(h-x)^2}{2u^2} } \right] u }{\sqrt{2\pi} (h-x)^2} .
\end{equation}
It is worth noting that, after the marginalization eliminates the unknown standard deviation from $N(x|h,\sigma)$, this same result is obtained by assuming the generalised Birge-ratio model $\sigma_i = \lambda_1 u_i$, where the scale parameters $\lambda_i \ge 1$ are datum-specific and take the uncertainty underestimations into account. As expected being known only its lower bound, the variance of (\ref{sampling:2}) is infinite.

Eventually, by taking (\ref{sampling:2}) into account, the data likelihood (\ref{lik}) can be rewritten as
\begin{equation}\label{lik:2}
 L(\bx|h,M_A) = \prod_{i\in A, j\in \bar{A}} N(x_i|h,u_i) G(x_j|h,u_j) ,
\end{equation}
which is parameterised only by $h$, to which the pre-data distribution $\pi(h)=1/V_0$ corresponds.

\subsection{Results}
\begin{figure}\centering
\includegraphics[width=7.5cm]{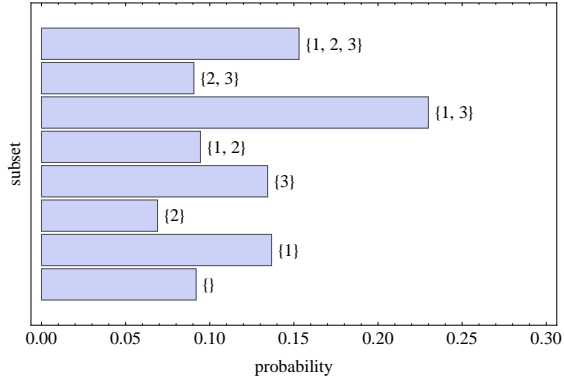}
\caption{Probabilities of the subsets of good data. The bar labels signify the good-data subset.} \label{figure:2}
\end{figure}

The three Planck constant values are grouped into eight $\{ A, \bar{A} \}$ subset-pairs, where the $A$ subsets collect the $\sigma_i = u_i$ good data and the $\bar{A}$s the $\sigma_i \ge u_i$  ones. The subset pairs are sorted according increasing cardinality of $A$, with $A = \varnothing$ first and later elements in table \ref{T1} omitted first. The $A$'s probabilities are given by (\ref{model}), where we assumed $\Pi(M_A)=$ const., the data evidence is
\begin{equation}\label{int}
 Z(\bx|M_A) = \frac{1}{V_0} \int_{-\infty}^{+\infty} L(\bx|h,M_A) \, {\rm d}h ,
\end{equation}
and $L(\bx|h,M_A)$ is given by (\ref{lik:2}). The integration in (\ref{int}) must be carried out numerically; the results are given in Fig.\ \ref{figure:2}.

All the subsets are roughly equally probable; none stands clearly out. The probability that at least one of the uncertainty values was underestimated is 85\%; conversely, the probability of a purely statistical origin of the data scatter is 15\%. The probabilities of the good-data subsets including the NIST value -- the second in the data list -- are local minima. Additionally, the most probable good-data set excludes it; this might suggest that the uncertainty associated to the NIST value is underestimated.

\begin{figure}\centering
\includegraphics[width=7.5cm]{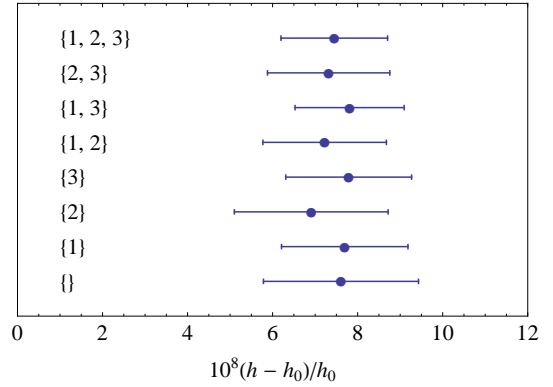}
\caption{Mean values and standard deviations of the post-data probability distribution of the possible values of $h$ given the values labelled 1 -- 3 in Table \ref{T1} and the indicated subsets of good data. $h_0 = 6.62606957\times 10^{-34}$ J s is the value recommended by the Committee on Data for Science and Technology.} \label{figure:3}
\end{figure}

The mean values and standard deviations of the post-data probability distributions of the possible values of $h$ for any subset of good data are shown in Fig. \ref{figure:3}. None value differs significantly from the others. It is also worth noting that the scale factor between the minimum standard deviation -- corresponding to the set including where all the data -- and maximum one -- corresponding to the empty set -- is 1.5.

\section{Fixed effect model}
In order to check the data consistency, we can also compare the assumption that the data are sampled or not sampled from distributions having the same mean. In this case, the hypotheses are as follows: The measured values are independently sampled from distributions having the same mean (H0, null hypothesis) or different means (H1, alternative hypothesis). In both cases, the standard deviations are equal to the associated uncertainties.

As before, this information is synthesised by Gaussian sampling distributions; the data likelihoods are
\begin{subequations}\begin{equation}
 L(\bx|h,\rH 0) = \prod_i N(x_i|h,u_i) ,
\end{equation}
which is parameterized by the one-dimensional coordinate $h$, and
\begin{equation}
 L(\bx|\bh,\rH 1) = \prod_i N(x_i|h_i,u_i) ,
\end{equation}\end{subequations}
which is parameterized by the three-dimensional coordinate $\bh=[h_1,h_2,h_3]^{\rm T}$.

With uniform probability densities in the H0 and H1 hypothesis-spaces, the pre-data distributions of the space coordinates are $\pi(h)=1/V_0$ (H0 hypothesis) and $\pi(\bh)=1/V_0^3$ (H1 hypothesis), where the volumes $V_0$ and $V_0^3$ of the $h$ and $\bh$ spaces are large enough to allow the limits of the evidence integrals to be extended up to the infinity.

Eventually, the data evidences are given by
\begin{subequations}\begin{equation}\label{Z01}
 Z_0 = \frac{1}{V'_0} \int_{V'_0} L(\bx'|h',\rH 0)\, \rmd h' = \frac{1.6\times 10^{-3}}{V'_0} ,
\end{equation}
and
\begin{equation}
 Z_1 = \frac{1}{V_0^{'3}} \int_{V_0^{'3}} L(\bx'|\bh',\rH 1)\, \rmd h'_1 \rmd h'_2 \rmd h'_3 = \left( \frac{1}{V'_0} \right)^3 ,
\end{equation}\end{subequations}
where the dimensionless variables $x' = 10^8x/h_0$, $h' = 10^8 h/h_0$, $u' = 10^8 u/h_0$, and $V'_0 = 10^8 V_0/h_0$ were used.

The inverse proportionality of (\ref{Z01}-b) to the volumes of the hypothesis-spaces is known as Ockham's razor and penalizes the model having the greater adaptability to the data. Consequently, in the case of a large pre-data range of the $h$ values, that is, when $V'_0 \rightarrow \infty$, the data support always the consistent-data hypothesis. Since -- in order to extend the limits of the (\ref{Z01}-b) integrals to the infinity -- we assumed that $V_0 \gg \max(u_1,u_2,u_3)$, $V'_0$ must be at least 10. In this case, $Z_0= 1.6\times 10^{-4}$, $Z_1= 10^{-3}$, and the probability of consistent data is about 16\%; not far from the value found in the previous analysis.

\begin{figure}\centering
\includegraphics[width=7.5cm]{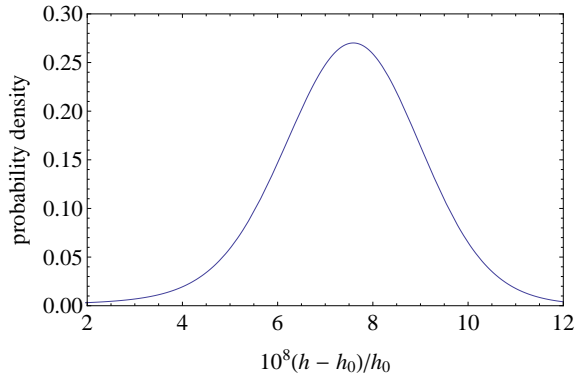}
\caption{Probability density of the possible values of the Planck constant given the values labelled 1 -- 3 in table \ref{T1} whatever the subset of good data may be. $h_0 = 6.62606957\times 10^{-34}$ J s is the value recommended by the Committee on Data for Science and Technology.} \label{figure:4}
\end{figure}

\section{Conclusions}
To implement a kilogram definition based on the Planck constant they are necessary at least three measured values having uncertainties less than $5\times 10^{-8} h$, with at least a value having an uncertainty less than $2\times 10^{-8} h$. These values must be consistent at the 95\% level of confidence. In addition, the  Task Group on Fundamental Constants of the Committee on Data for Science and Technology must provide a value that minimises discontinuities.

This paper investigated the data consistency by explaining the measurement results by random effect -- which allow, but not assume, missing contributions to the error budgets -- and fixed effect -- which allow, but not assume, different means of the sampling distributions -- models. In both cases, the data look inconsistent.

In the first case, after averaging over all the subsets of good data, the probability density of the possible values of $h$ is shown in Fig.\ \ref{figure:4}. The mean value is
\begin{equation}\label{hmean}
 h = 6.626070073(94) \times 10^{-34} \; \rm J s .
\end{equation}
The standard deviation, $1.4\times 10^{-8} h$, can be compared with the weighed-mean uncertainty, $1.3\times 10^{-8} h$. The quadratic difference, $0.5\times 10^{-8} h$, is the contribution to the error budget of the uncertainty about the actual subset of good data.

\section*{Acknowledgements}
This work was jointly funded by the European Me\-trology Research Pro\-gramme (EMRP) par\-ti\-ci\-pa\-ting coun\-tries within the European Association of National Metrology Institutes (EURAMET), the European Union, and the Italian ministry of education, university, and research (awarded project P6-2013, implementation of the new SI).





\end{document}